\documentclass[preprint,12pt]{elsarticle}
\usepackage{amssymb}
\usepackage{tabularx}
\usepackage[export]{adjustbox}
\usepackage{booktabs}
\usepackage[vlined, boxed]{algorithm2e}
\usepackage{caption}
\journal{Journal of Nuclear Materials}
 \usepackage[dvipsnames]{xcolor}
 \usepackage{tikz, tikz-3dplot}
\usetikzlibrary{positioning, calc, shapes, chains, arrows}
\usetikzlibrary{decorations.pathreplacing,calligraphy}
\tikzset{
  font={\fontsize{9pt}{12}\selectfont}}
 \usepackage{amsmath} 
 \usepackage{subcaption}
 \usepackage{comment}
 \usepackage{tikzscale}
\usepackage{float}
\usepackage{pgfplots}
\DeclareUnicodeCharacter{2212}{−}
\usepgfplotslibrary{groupplots,dateplot}
\usetikzlibrary{patterns,shapes.arrows}
\pgfplotsset{compat=newest}
\begin{document}
\definecolor{pythonBlue}{HTML}{1f77b4}
\definecolor{pythonOrange}{HTML}{ff7f0e}
\begin{frontmatter}

\title{\textcolor{black}{
A new Surrogate Microstructure Generator for Porous Materials with Applications to
the Buffer Layer of TRISO Nuclear Fuel Particles}}
\author[inst1]{Philipp Eisenhardt}
\author[inst2]{Ustim Khristenko}
\author[inst3]{Barbara Wohlmuth}
\author[inst1]{Andrei Constantinescu}
     
\affiliation[inst1]{organization={Laboratoire de Mécanique des Solides, Ecole Polytechnique, CNRS, Institut Polytechnique de Paris},
            postcode={91128},
            city={Palaiseau},
            country={France}}

\affiliation[inst2]{organization={Safran Tech, Digital Sciences \& Technologies Department},
            city={Châteaufort},
            postcode={78114}, 
            state={Magny-Les-Hameaux},
            country={France}}
       
\affiliation[inst3]{organization={Department of Mathematics,Technical University of Munich}, 
            postcode={85748}, 
            city={Garching b. München},
            country={Germany}}

\begin{abstract}
We present a surrogate material model for generating microstructure samples reproducing the morphology of the real material.
The generator is based on Gaussian random fields, with a Matérn kernel and a topological support field defined through ellipsoidal inclusions clustered by a random walk algorithm. 
We identify the surrogate model parameters by minimizing misfits in a list of statistical and geometrical descriptors of the material microstructure. 
To demonstrate the effectiveness of the method for porous nuclear materials, we apply the generator to the buffer layer of Tristructural Isotropic Nuclear Fuel (TRISO) particles. 
This part has been shown to be failure sensitive part of TRISO nuclear fuel and our generator is optimized with respect to a dataset of FIB-SEM tomography across the buffer layer thickness.
We evaluate the performance by applying mechanical modeling with problems of linear elastic homogenization and linear elastic brittle fracture material properties and comparing the behaviour of the dataset microstructure and the surrogate microstructure.
This shows good agreement between the dataset microstructure and the generated microstructures over a large range of porosities. 
\end{abstract}
\begin{graphicalabstract}
\includegraphics[width=\linewidth]{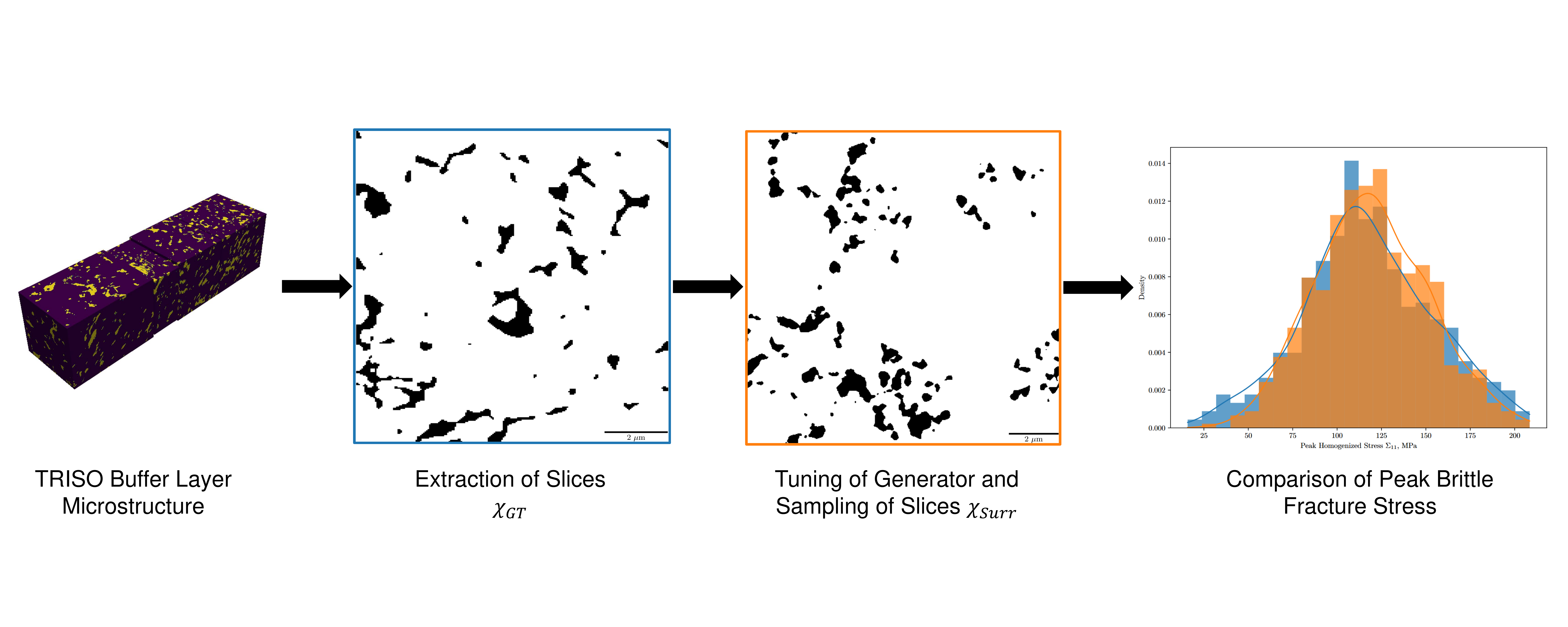}
\end{graphicalabstract}

\begin{highlights}
\item This paper proposes a novel generator of surrogate models of material microstructure including randomness.
\item The surrogate model is constructed using level-set functions, Gaussian random fields with Matérn Kernel and stochastic optimisation. 
\item The parameters of the surrogate model are obtained by minimizing various distances between model and real data.
\item The effectiveness of the model is illustrated on a porous material, respectively the buffer layer of the TRISO nuclear fuel particles.
\item The estimated mechanical properties of the surrogate material (elastic moduli, damage behavior, etc) are in good agreement with estimations from the real material over a range of global porosities.   
\end{highlights}
\end{frontmatter}

\section{Introduction}
\label{sec:Intro}
To meet the evolving demands of modern nuclear reactors, Tristructural Isotropic (TRISO) nuclear fuel particles have emerged as a key technology in various countries for multiple reactor designs \cite{powers_review_2010}. TRISO fuel is a central focus of the U.S. Department of Energy's Advanced Gas Reactor (AGR) Fuel Development and Qualification Program, which aims to optimize and qualify this advanced fuel technology \cite{demkowicz_triso_2021, demkowicz_doe_2023}.
TRISO fuels offer significant advantages over traditional large uranium fuels, particularly in their ability to be adapted to varying power demands and their enhanced safety profile. Their small size not only makes fueling more flexible but also reduces the risk of catastrophic failure, as the containment of radioactive materials is more robust.
\goodbreak

A TRISO particle consists of a central fuel kernel, which contains the radioactive material where nuclear fission occurs. The fuel kernel is surrounded by a porous carbon buffer layer that absorbs the expansion of the fuel during irradiation and accommodates fission gases. This buffer layer, which is the primary focus of this paper, plays a crucial role in managing the internal stresses generated during reactor operation.
The buffer layer is followed by the Inner Pyrolytic Carbon (IPyC) layer, which retains fission products and enhances the particle's mechanical integrity. Surrounding the IPyC layer is a silicon carbide (SiC) layer, a robust ceramic material that provides an impermeable barrier against the release of fission products. The final layer is the Outer Pyrolytic Carbon (OPyC) layer, which further protects the SiC layer from external damage and supports the structural integrity of the particle.
We refer to \cite{condon_fate_2021, demkowicz_triso_2021} and the references therein for additional details. 
\goodbreak
A large review of their performance and mechanical failure has been conducted in \cite{powers_review_2010}, however, the description is based on empirical equations and the behaviour depends purely on the orientation distribution and porosity. 
Whilst failure of the TRISO particles is rare, its onset usually stems from failure of the buffer carbon, where circumfential tearing is the most frequent failure mode which can lead to fracture of IPyC layer and the SiC layer through the exposition of fission products and corrosion \cite{seibert_agr-2_2023, gerczak_role_2020, hunn_initial_2018}. 
To model the TRISO particle performance, the PARFUME and BISON framework \cite{skerjanc_parfume_2018,jiang_triso_2021} can be used, however, the different layers of the TRISO particles are modeled to be homgeneous and a focus is on the multiphysics coupling and the behaviour of the entire fuel particle. 
\goodbreak
The microstructure of a material plays a critical role in determining its mechanical and physical properties, making its characterization a central challenge in materials science. Local deviations in microstructure influence the onset of plasticity, brittle or ductile fracture, fatigue, and also affect the material's global properties and other phenomena.

In \cite{griesbach_microstructural_2023}, the buffer layer microstructure of Tristructural Isotropic Nuclear fuel particles was characterized and studied through focused ion beam scanning electron microscopy (FIB SEM). 
Here, FIB SEM tomography was performed throughout the buffer layer of surrogate TRISO particles, where the uranium kernels was replaced by a zirconia kernel. 
The importance of circumfential tearing and potential sources of this failure mode were studied and an extensive study on the buffer layer pores was conducted. 
While past research has extensively examined the geometric characterization of the microstructure, particularly pore distribution in the buffer layer \cite{griesbach_microstructural_2023}, and the fracture behavior of the buffer under irradiation-induced stress \cite{hosemann_mechanical_2013, rohbeck_evaluation_2016, griesbach_irradiation-condition_2025}, the coupled effects of pore structure and fracture dynamics remain largely unstudied. 
Understanding this coupling is essential for improving the performance and safety of TRISO fuel under operational conditions, as failure in the buffer layer could compromise the integrity of the entire fuel particle. Recent studies \cite{recuero_fracture_2024} highlight the importance of developing more comprehensive models that account for both the microstructural and mechanical behavior of TRISO particles.

To generate surrogate samples from limited, measured microstructural data, different proposals and methods have been made. 
An overview can be can be found in \cite{torquato_random_2002, bargmann_generation_2018,jeulin_morphological_2021}. 
For the give type of porous microstructure, four different methods are of particular interest. 
In \cite{chakraborty_phase-field_2016}, a packing based approach was studied to model microstructure dependent fracture in nuclear grade materials. 
Here, the microstructure is generated by packing spherical pores leading to matched size shape distributions of the surrogate microstructure, however, at the cost of a lack of variability. 
Furthermore, methods based on Gaussian random fields as described in \cite{torquato_random_2002, robertson_efficient_2022} can be applied. 
Here, two point statistics as the two point correlation function will be matched, however, deviations in the size distribution functions of the pores can arise. 
In order to match these statistics, approaches based on energy minimization principles have been developed in \cite{yeong_reconstructing_1998, cule_generating_1999, seibert_agr-2_2023}, however, here the optimization needs to be repeated for each additional microstructure sample.
As a hybrid approach, the combination of packing based methods and Gaussian random field based methods has been developed in \cite{khristenko_statistical_2020, khristenko_statistically_2022}, where the resulting surrogate microstructure is generated based on the weighted sum of a topological support field and Gaussian random field. 
This approach promises to be capable of the inexpensive generation of additional surrogate samples, however, requires significant fine tuning and prior knowledge. 
Recently, approaches including Deep Learning approaches such as Generative Adversarial Networks (GANs) studied as in \cite{li_transfer_2018, henkes_three-dimensional_2022}, where the optimization is conducted by minimizing the Gram misfit. 
Whilst these approaches provide well adapted microstructures for a wide range of applications without a large amount of subject specific adaptations, they require more data and the resulting parameters are difficult to interpret. 
This work builds onto the hybrid approach with Gaussian Random Field and underlying topological support introduced in \cite{khristenko_statistical_2020, khristenko_statistically_2022}, however, the definition of the topological support has been adapted to tune it to the application of the TRISO buffer layer. 
Additionally, the identification method has been extended using the method of efficient global optimization (EGO) to identify the parameters \cite{jones_efficient_1998}. 
\goodbreak
In Section \ref{sec:surMSGenId} the surrogate microstructure generator is introduced and Section \ref{sec:surrApplDef} details microstructure descriptors and their computation and the identification of surrogate microstructure parameters based upon this. 
Section \ref{sec:mechProb} introduces the mechanical problem the dataset microstructure and surrogate samples are subjected to. 
In Section \ref{sec:TrisoProb}, the application of the introduced method to the buffer layer of the TRISO microstructure from \cite{griesbach_microstructural_2023} is introduced. 
The results are presented in Section \ref{sec:res} and discussed with conclusions in \ref{sec:concs}.
\section{Surrogate Microstructure Method}
\label{sec:surMSGenId}
Let us consider the material domain $\mathcal{D} \subset \mathbb{R}^n$ and a set of material distributions in $\mathcal{D}$ indexed by a real number $\omega$. 
For a two-phase material, the set of microstructures will be described using the characteristic function of one of the phases, denoted as $\chi_{D}[\omega](\boldsymbol{x})$. The notation indicates the fact that the material phase $\chi_{D}$ is a function of both the spatial position $\boldsymbol{x}$ in the domain $\mathcal{D}$ and each sample $\omega$. 

The corresponding numerically created surrogate microstructure will equally be expressed by a characteristic function of the material phase, denoted as $\chi_{S}[\omega, \theta](\boldsymbol{x})$. One can remark that the surrogate microstructure depends on an additional argument, $\theta$, the parameter vector defining the generator of the surrogate microstructure. Hence, the surrogate microstructure depends on the position $\boldsymbol{x}$, $\theta$ the parameter vector of the generator and a sample indicator $\omega$. 

The characteristic function of the material and surrogate model, $\chi_{D}[\omega](\boldsymbol{x})$ and $\chi_{S}[\omega, \theta](\boldsymbol{x})$ respectively will later be defined as cuts of level set functions. We further assume that $\chi_{D}[\omega](\boldsymbol{x})$ is known for several samples $\omega$ by direct measurements and observations of the real material and that $\chi_{S}[\omega, \theta](\boldsymbol{x})$ will be numerically be constructed using the observed data. 

Moreover, the vector of optimal parameters $\theta^*$ of the surrogate generator $\chi_{S}$ is obtained as the solution of the following stochastic optimization problem: 
\begin{equation}
	\theta^*=\text{argmin}_\theta \left( \sum_{i}^{N_{d}} \alpha_i d_i(p_i\left(\chi_{D}\right), p_i\left(\chi_{S}(\theta)\right)) \right), 
	\label{eq:genCostFun}
\end{equation}
where the cost function to be minimized is a weighted sum of the distances $d_i$, measuring deviations between the real and surrogate samples. 
$p_i$ denotes the descriptor and $i$ and $N_{d}$ is the index of the image descriptor, where each descriptor is weighted by $\alpha_i$.

As detailed next, the distances can either express global descriptors, such as the correlation functions or the lineal path function, or image local image descriptors, such as pore area, size, ellipticity or solidity in the case of a porous material. 

\textit{Surrogate Microstructure Model.} The construction of the surrogate microstructure $\chi_{S}[\omega, \theta](\boldsymbol{x})$ is based on the work in \cite{khristenko_statistical_2020, khristenko_statistically_2022-1}. In the case of a porous microstructure, which corresponds to the example discussed in this paper, it constructs the characteristic function of the microstructure $\chi_{S}$ by adding a smooth \textit{topological support}, i.e. a level-set description of particles or pores $\phi_{\text{T}}$ and a noisy \textit{perturbation field}, i.e. a Gaussian Random Field (GRF) $\phi_{\text{GRF}}$, creating a randomly perturbed morphology. 

\textit{Level Sets} Both fields $\phi_{\text{T}}$ and $\phi_{\text{GRF}}$, i.e. the topological support and the Gaussian Random Field respectively, are constructed through cuts from level set functions as proposed in \cite{osher_fronts_1988}. 

In the case of a porous material described by two phases: the pores and a matrix, the morphology is defined by the characteristic function $\chi_{\text{S}}$ by the cut of the global level set function $\phi_{\text{S}}[\omega, \theta](\boldsymbol{x})$ in the following way:
\begin{align}
	\chi_{\text{S}}[\omega, \theta](\boldsymbol{x})=
	\begin{cases}
		1, & \text{if } \phi_{\text{S}}[\omega, \theta](\boldsymbol{x})<\tau, \text{ pore} \\
		0, & \text{if } \phi_{\text{S}}[\omega, \theta](\boldsymbol{x})\geq\tau, \text{ matrix}, 
	\end{cases}
\end{align}
Let us remark that the cutoff parameter $\tau$ implicitly defines the volume fraction $\varphi_s(\omega)$ for given $\chi_s(\omega)$ 
\begin{align}
	\label{eq:vfLS}
	\varphi(\omega)=\frac{\int_\mathcal{D} \chi_S(\boldsymbol{x}) \text{d}V}{\int_\mathcal{D} 1\text{d}V}. 
\end{align}
The preceding formula holds true for all level sets. 

\paragraph{Level Set Mixture}
The surrogate microstructure is defined as the weighted combination of the two level sets:
\begin{align*}
	\phi_{\text{S}}[\omega, \theta](\boldsymbol{x})=(1-\alpha)\phi_{\text{T}}[\omega, \theta_{T}](\boldsymbol{x})+\alpha\phi_{\text{GRF}}[\omega, \theta_{GRF}](\boldsymbol{x}), 
\end{align*}
where the noise level $\alpha$ is part of the surrogate generator's parameters $\theta_{GRF}$. 
The noise level can take values between 0 and 1, where 0 leads to a microstructure purely defined by the topological support and 1 leads to realizations without topological support. 

\paragraph{Topological Support - Ellipsoid  Definition}
We shall describe the construction of the level set function for ellipsoidal inclusions with a random walk algorithm. 
The topological support is constructed from a series of ellipsoidal inclusions with centers using a random walk algorithm. 
The level set of each particle is computed as
\begin{equation}
	\phi_{\text{T}}^i(\boldsymbol{x})=\lVert \boldsymbol{x}-\boldsymbol{x}^i\lVert_{\boldsymbol{A}, \beta_i}^2= (\boldsymbol{x}-\boldsymbol{x}^i)\boldsymbol{Q}^T(\beta_i)\boldsymbol{A}\boldsymbol{Q}(\beta_i)(\boldsymbol{x}-\boldsymbol{x}^i),
\end{equation}
where the distribution of the centers $\boldsymbol{x}^i$, axis lengths $\boldsymbol{a}^i$ and orientations $\beta_i$ are part of the parameter vector $\theta_T$. 
$\boldsymbol{Q}(\beta_i)$ denotes the rotation matrix by angle $\beta_i$ and $\boldsymbol{A}$ denotes the axis length matrix from the individual axis lengths $a_i$. 
This leads to a level set for each ellipse $e_i$, where the resulting level set is the minimum of the individual level sets of all the $N_e$ ellipsoids. 
For further details, see \cite{khristenko_statistically_2022-1}. 
This method is explained in algorithm \ref{alg:elDef}.

\paragraph{Topological Support - Ellipsoid Clustering}
To ensure larger inclusions with complex morphologies, ellipsoids will be clustered in larger pores using a random walk algorithm to define the position of their centers. 
The number of random walk seeds $N_c$ is defined through a binomial distribution 
\begin{equation*}
	N_c=B(N_c^*, p_c^*),
\end{equation*}
meaning that the number of cluster seeds is sampled from a binomial distribution with parameter $N_c^*$ and a success probability of $p_c^*$. 
$N_c^*$ is part of the parameter vector $\theta_T$ and will be optimized.
For each of the random walk seeds, the starting point $\boldsymbol{x}^{i0}$ is computed based on a uniform distribution on the numerical domain to obtain $\boldsymbol{x}^{i0}\sim\mathcal{U}(\mathcal{D})$.
Given an already sampled center $\boldsymbol{x}_e^{i(j-1)}$, we define the next center  $\boldsymbol{x}_e^{ij}$ using the following formula
\begin{align}
	\boldsymbol{x}_e^{ij}\gets \boldsymbol{x}_e^{i(j-1)}+r\sin{\left(\alpha^{ij}\right)}\boldsymbol{e}_x+r\cos{\left(\alpha^{ij}\right)}\boldsymbol{e}_y, 
\label{eq:randomWalkSteps}
\end{align}
where parameters of the random walk length $r$ and direction $\alpha^{ij}$ are part of the parameter set $\theta_T$.
The random walk starts from a seed point $\boldsymbol{x}^{i0}\sim \mathcal{U}(\mathcal{D})$, which is sampled from a uniform distribution on the numerical domain $\mathcal{D}$. 
To compute the resulting level set from the individual particle level sets $\phi_P^k$, the minimum needs to be computed leading to the topological support level set
\begin{equation}
    \phi_T[\omega, \theta](\boldsymbol{x})=\min_k\left(\phi_P^k[\omega, \theta](\boldsymbol{x})\right).
\end{equation}
\begin{figure}
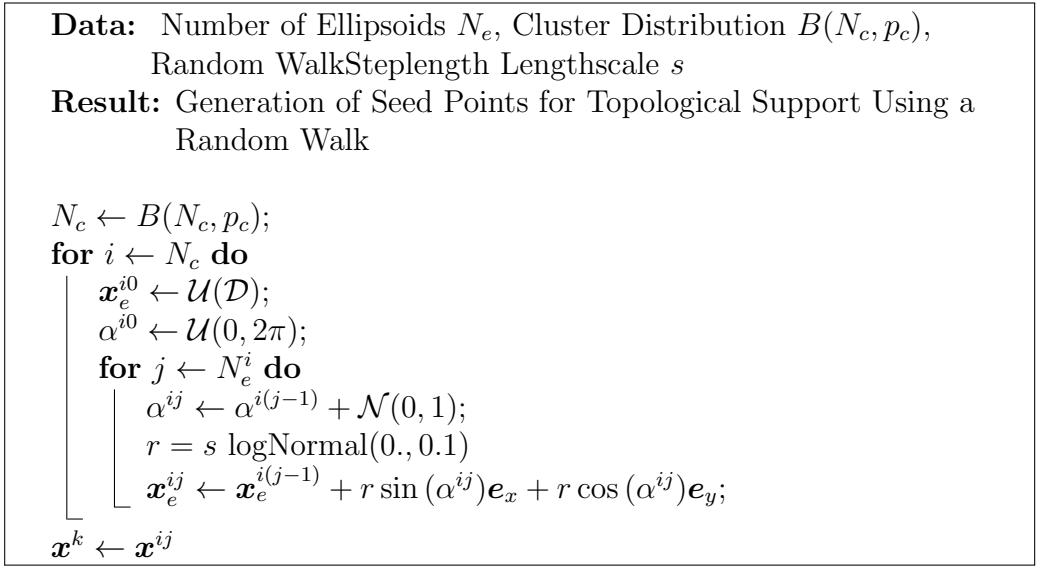

\begin{algorithm}[H]
	\KwData{
		Number of Ellipsoids $N_e$,
		Cluster Distribution $B(N_c, p_c)$, 
		 Random WalkSteplength Lengthscale $s$}
	\KwResult{Generation of Seed Points for Topological Support Using a Random Walk}
\vspace{0.5cm}
    $N_c\gets B(N_c, p_c)$;\\
	\For{$i\gets N_c$}{
		$\boldsymbol{x}_e^{i0}\gets \mathcal{U}(\mathcal{D})$;\\
		$\alpha^{i0}\gets \mathcal{U}(0, 2\pi)$;\\
		\For{$j\gets N_e^i$}{
			$\alpha^{ij}\gets \alpha^{i(j-1)}+\mathcal{N}(0, 1)$;\\
			$r = s \ \text{logNormal}(0., 0.1)$\\
			$\boldsymbol{x}_e^{ij}\gets \boldsymbol{x}_e^{i(j-1)}+r\sin{\left(\alpha^{ij}\right)}\boldsymbol{e}_x+r\cos{\left(\alpha^{ij}\right)}\boldsymbol{e}_y$;\\
		}
	}
    $\boldsymbol{x}^k \gets \boldsymbol{x}^{ij}$
\end{algorithm}
\caption{Algorithm describing the generation of clustered ellipsoidal inclusions using a Random Walk based algorithm. The method gives rise to the ellipsoid centers $\boldsymbol{x}^{ij}$.}
\label{alg:rwIncDef}
\end{figure}
\begin{figure}
\begin{algorithm}[H]
	\KwData{
		Ellipsoid Centers $x_c$, 
		Maximum Aspect Ratio $q_{max}$, 
	}
	\KwResult{Level set of the topological support}
\vspace{0.5cm}
	\For{$k\gets N_e$}{
		$q_{max}\gets r q_{min}$;\\
		$d_1^k\gets \mathcal{U}[q_{min}, q_{max}]$;\\
		$d_2^k\gets \mathcal{U}[q_{min}, q_{max}]$;\\
		$\mathbf{D}\gets 
		\begin{pmatrix}
			1/d_1^k & 0\\
			0 & 1/d_2^k
		\end{pmatrix}$;\\
		$\beta^k \gets \mathcal{U}[0, 2 \pi]$;\\
		$\mathbf{A} \gets \mathbf{Q}^T(\beta^k) \mathbf{D}\mathbf{Q}(\beta ^k)$;\\
		$\phi_k(\boldsymbol{x}) \gets (\boldsymbol{x}-\boldsymbol{x}_e^k)\boldsymbol{A}(\boldsymbol{x}-\boldsymbol{x}_e^k)$;\\
	}
	$\phi_{\text{T}}(\boldsymbol{x}) \gets \min_k{\phi_k(\boldsymbol{x})}$
\end{algorithm}    
\caption{Algorithm illustrating the computation of the topological support level set based on the individual ellipsoid's properties. The algorithm requires the computed ellipsoid centers computed as in Algorithm \ref{alg:rwIncDef} and subsequently computes the level set of the topological support. }
\label{alg:elDef}
\end{figure}
\begin{figure}[h]
	\centering
	\includegraphics[width=\columnwidth]{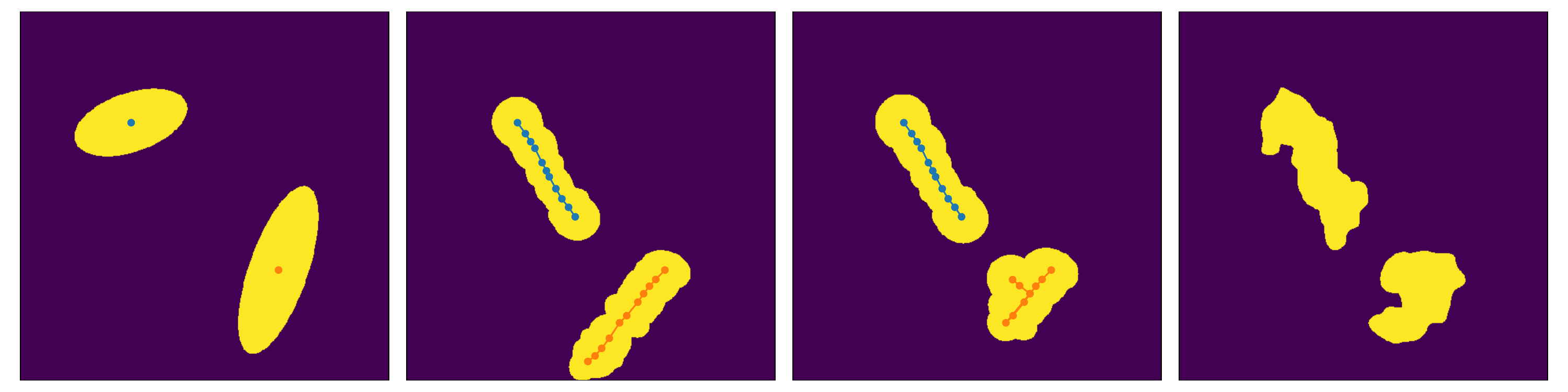}
\begin{tabularx}{\linewidth}{*{4}{>{\centering\arraybackslash}X}}
(a) & (b) & (c) & (d)  \\
    \vspace{-.4cm}
    \end{tabularx}
    \vspace{-.7cm}
        \caption{Visualization of the level set and surrogate inclusion generation for different parameters. a): Ellipsoid based on the chosen aspect ratios and orientations (see \ref{alg:elDef}), b): 12 ellipsoidal inclusions with centers based on Random Walk Algorithm (see \ref{alg:rwIncDef}), c): Inclusions generated from random walk algorithm with forking for one of the clusters.  
    d):Generated Inclusions with Distortions based on GRF }
	\label{fig:partSampStruct}
\end{figure}
\paragraph{Random Walk Fork}
To obtain large pore clusters with concave shapes, a forking method is introduced. 
Here, with a probability $p_s$, the random walk shows a fork, where after half of the random walk steps, a fork at a $90^\circ$ angle is introduced. 
Thus, Equation \ref{eq:randomWalkSteps} needs to be adapted to account for this change. 
In practice, this leads to the following equation with 
\begin{align}
	\boldsymbol{x}_e^{ij}&\gets \boldsymbol{x}_e^{i(j-1)}+r\sin{\left(\alpha^{ij}\right)}\boldsymbol{e}_x+r\cos{\left(\alpha^{ij}\right)}\boldsymbol{e}_y, &\text{for } j<j^*\\
	\boldsymbol{x}_e^{ij}&\gets \boldsymbol{x}_e^{i(j-1)}+r\sin{\left(\alpha^{ij}+90^\circ\right)}\boldsymbol{e}_x+r\cos{\left(\alpha^{ij}+90^\circ\right)}\boldsymbol{e}_y, &\text{for } j=j^*\\ 
    \boldsymbol{x}_e^{ij}&\gets \boldsymbol{x}_e^{i(j-1)}+r\sin{\left(\alpha^{ij}\right)}\boldsymbol{e}_x+r\cos{\left(\alpha^{ij}\right)}\boldsymbol{e}_y, &\text{for } j>j^*, 
\label{eq:randomWalkFork}
\end{align}
which means that after half of the ellipse centers have been computed, the forking occurs with half of the remaining ellipses being added as previously in Equation \ref{eq:randomWalkSteps} and the other half is added based on a 90 degree shift. 
This ensures more concave shapes. 
The forking can be generalized by inferring the forking points as a further parameter or having multiple forks, however, in the context of the given problem this was omitted. 
\paragraph{Noise Perturbation Field Generation}
The noise level set $\phi_{\text{GRF}}[\theta, \omega]\ (boldsymbol{x})$ is sampled through a Gaussian Random Field with a homogeneous covariance function. 
Let us consider the following Matérn type covariance function \cite{whittle_stochastic-processes_1963}
\begin{align}
	\label{eq:matKern}
	\mathcal{C}(\boldsymbol{x}, \boldsymbol{y})=\mathcal{M}_\nu\left(\frac{\sqrt{2\nu}}{l}\lVert \boldsymbol{x}-\boldsymbol{y} \lVert_{\boldsymbol{A}}\right), &&
	\mathcal{M}_\nu(\boldsymbol{x})=\frac{\boldsymbol{x}^\nu \mathcal{K}_\nu(\boldsymbol{x})}{2^{\nu-1}\Gamma(\nu)}, 
\end{align}
with $\mathcal{M}_\nu$ as the normalized Matérn kernel, where $\mathcal{K_\nu}$ is the modified Bessel function of second kind and $\Gamma$ is the gamma function with parameters $\nu$ denoting the smoothness parameter defining how many times the resulting random field will be differentiable, and $l$ denoting the correlation length of the random field. 
This leads to a random field with Matérn type covariance, leading to a covariance depending on the separation between two points $\boldsymbol{x}, \boldsymbol{y}$.
Matérn type random fields have seen popularity in different engineering fields for the modeling of geometric and mechanical uncertainties (see for example \cite{khristenko_statistical_2020, zhang_representing_2024, patinet_quantitative_2013, koh_stochastic_2023, chu_stochastic_2021} and the references therein).
The correlation length $l$ and smoothness $\nu$ are numerical parameters to be inferred as part of $\theta_{GRF}$.  
Algorithmically, the generation follows the structure laid out in \cite{ravalec_fft_2000, khristenko_statistically_2022-1, khristenko_analysis_2019} using an FFT based approach to ensure fast generation times. 
Here, the sampling is conducted by computing the power spectrum of the covariance function.
This can be written as 
\begin{align}
    z\sim\mathcal{N}(0, 1), && Z=FFT\{z\},  && G=\sqrt{dx S}, && q=FFT^{-1}\{G\cdot Z\},
\end{align}
with $z$ being independent and identically distributed, $S$ as the power spectrum and $dx$ being derived from the gridsize, and $q$ is the normally distributed output Gaussian random field with Matérn covariance. 
For a covariance function following Equation \ref{eq:matKern} its Fourier transformation, which yields the power spectrum spectrum $S$, can be written as
\begin{equation}
    S(f)=\frac{2^n\pi^{n/2}\Gamma(\nu+\frac{n}2)(2\nu)^\nu}{\Gamma(\nu)l^{2\nu}}\left(\frac{2\nu}{l^2} + 4\pi^2f^2\right)^{-\left(\nu+\frac{n}2\right)}, 
\end{equation}
where $f$ denotes the frequency in Fourier domain. 
\paragraph{Level Set Cutoff}
As previously stated, the surrogate microstructure is described by a resulting level-set $\phi$ from $\phi_\text{T}$ and $\phi_\text{GRF}$, which can be cut to have a binary microstructure representation. 
In order to match the desired volume fraction $\varphi_i$ of the microstructure, the cutoff parameter $\tau$ needs to be determined. 
It can easily be seen from Equation \ref{eq:vfLS} that for a choice of $\tau=\infty$, the volume fraction will be 1 and for  $\tau=-\infty$, the volume fraction will be zero. 
To match the description of dataset microstructure, volume fractions from this microstructure are hence sampled. 
\paragraph{Binary Filtering}
To avoid a large number of very small inclusions, a binary filtering method is employed with an opening procedure (binary erosion followed by a dilation) \cite{jeulin_morphological_2021}.
The opening of binary microstructure $\chi_{in}$ with structuring element $D$ to compute $\chi_{out}$ can be written as 
\begin{equation}
	\chi_{out}=\chi_{in} \circ D=(\chi_{in}  \ominus D)\oplus D, 
\end{equation}
 where D corresponds to a disk as a structuring element, and $\ominus, \ \oplus$ denote the erosion and dilation, respectively. 
 This parameter is optimized as part of $\theta_{GRF}$
 This operation is applied onto the cut level set. 
 For more details on this morphological operation and other operators, see \cite{jeulin_morphological_2021} and the references therein. 

\section{Application to Microstructural Data}
\label{sec:surrApplDef}
In order to apply to the surrogate model, the defined parameters $\theta$ need to be inferred. 
Hence, descriptors for microstructures are defined in Section \ref{sec:msDesc} and the definition of measures of similarity between different microstructures and the identification of parameters is introduced in Section \ref{sec:surrId}.

\subsection{Microstructure Description}
\label{sec:msDesc}
Two different types of microstructure descriptors were used for the course of this studies. 
The first set of descriptors focuses on global behavior at larger length scales and is based on slices from both the dataset and the surrogate microstructure. 
For this purpose, the Two-Point Correlation Function and the Lineal Path Function were selected.
The second set of descriptors captures local properties, calculated after segmenting the microstructures to isolate individual pores. From the segmented images, the area of each pore ($A_p$) and the area of its convex hull ($A_{CH}$) were computed. The convex hull of a pore is defined as the smallest convex shape that fully encloses it. 
A wide range of other descriptors can readily be used depending on the problem, see \cite{jeulin_morphological_2021, torquato_random_2002, van_der_walt_scikit-image_2014, igathinathane_shape_2008} and the references therein for other descriptors. 
Their inclusion into the optimization algorithm does not pose any inherent difficulties. 
\goodbreak
When comparing based on multiple slices, these global descriptors can be averaged following
\begin{equation*}
	\left<f(\boldsymbol{r}, \omega_i)\right>_\omega=\frac{1}{N_\omega} \sum_{i}^{N_\omega} f(\boldsymbol{r}, \omega_i), 
\end{equation*}
when working with different slices $\omega_ia$. 
\paragraph{Two Point Correlation Function} The two point correlation function $S_2$ of a microstructure is defined as the probability that two points connected by line segment $\boldsymbol{r}$ are both in the pore phase, leading to 
	\begin{equation}
		S_2(\boldsymbol{r})=\left<\frac{1}{V}\int_V \chi(\boldsymbol{x})\chi(\boldsymbol{x}+\boldsymbol{r}) \text{d}\boldsymbol{x} \right>_\omega, 
		\label{eq:defS2}
	\end{equation}
	which is the description for the ergodic case, where the descriptor only depends on the vector between the two points $\boldsymbol{r}$ and not on the points themselves ($\boldsymbol{x}_1$ and $\boldsymbol{x}_1+\boldsymbol{r}$).
	The extreme values of the two point correlation function are 
	\begin{align*}
		\lim_{\boldsymbol{r}\rightarrow\infty} S_2(\boldsymbol{r})=\varphi && \lim_{\boldsymbol{r}\rightarrow\infty} S_2(\boldsymbol{r})=\varphi^2
	\end{align*} 
	More details can be found in \cite{torquato_random_2002}.
\paragraph{Lineal Path Function} The lineal path function quantifies the probability that a line segment $\boldsymbol{r}$  fully lies in the pore phase \cite{lu_lineal-path_1992, torquato_random_2002}.
	\begin{equation}
		L (\boldsymbol{r} ) = \left<P
		\big\{
		\chi(\boldsymbol{x}, \omega_i) = 1 ,
		\chi(\boldsymbol{x}+\delta\boldsymbol{x}_1, \omega_i) = 1 ,
		\ldots
		\chi(\boldsymbol{x}+\boldsymbol{r}, \omega_i) = 1
		\big\}\right>_\omega, 
		\label{eq:defLp}
	\end{equation}
	which is again for the special case of an ergodic material. 
	The extreme values of the lineal path function are 
	\begin{align*}
		\lim_{\boldsymbol{r}\rightarrow\infty} L_p(\boldsymbol{r})=\varphi && \lim_{\boldsymbol{r}\rightarrow\infty} L_p(\boldsymbol{r})=0.
	\end{align*}
	For the algorithms, the implementation from \cite{de_geus_tdegeusgooseeye_2024} was used.
	\paragraph{Size} The most simple descriptor of a pore is its size. 
	Here, the size of the segmented pore is computed, and this process is repeated for all pores in the ensemble. 
	\begin{equation}
		I_{Si}^i=A_p^i
		\label{eq:defSi}
	\end{equation}
	If the pore is convex, this descriptor will be 1; for very concave pores, the descriptor will approach 0. 
	For all the pores in the ensemble $\omega$ this descriptor is computed. 
	See \cite{van_der_walt_scikit-image_2014} for further details on the implementation. 	
	\paragraph{Solidity} A pore's solidity is defined as the ratio of the pore area to the area of its convex hull 
	\begin{equation}
		I_{So}^i=\frac{A_p^i}{A_{CH}^i}.
		\label{eq:defSo} 
	\end{equation}
	If the pore is convex, this descriptor will be 1, for very concave pores the descriptor will approach 0. 
	For all the pores in the ensemble $\omega$ this descriptor is computed. 
	See \cite{van_der_walt_scikit-image_2014} for further details and the corresponding scikit-image documentation. 

\subsection{Surrogate Generator Parameter Identification}
\label{sec:surrId}
As previously noted, in order to get an appropriate surrogate microstructure generator, the parameters $\theta$ need to be identified based on a cost function as introduced in equation \ref{eq:genCostFun}. 
The general cost function was specified using classical microstructure descriptors as introduced in \cite{torquato_random_2002}. 
The morphological descriptors applied in this study were the two point correlation function $S_2$ and the lineal path function $L_p$ and on an individual pore level, the size distribution function and the solidity distribution function, which are explained in Section \ref{sec:msDesc}.
\goodbreak 
In order to reduce the dimensionality of the problem, the optimization was split into two parts, where first, the parameters of the topological support are optimized ($\theta_{T}$), and in the second step, the parameters of the noise field ($\theta_{GRF}$). 
This is repeated in a staggered scheme, with the parameters that are not being optimized being kept fixed at the result from the previous iteration.
The two iteration parts are based on different cost functions. 
The topological support is tuned based on the cost function $J_{SS}$, which is written as 
\begin{equation}
	J_{SS}(\theta, \omega)= \beta D_{KL}(P_{Si}^*||P_{Si}(\theta, \omega)) +(1-\beta)D_{KL}(P_{So}^*||P_{So}(\theta, \omega)),
\end{equation}
and includes the Kullback-Leibler (KL) divergence $D_{KL}$ between the dataset microstructure's pore sizes or solidities and the surrogate microstructure's pore sizes or solidities, hence accounting for misfits on a pore-based level.
Hence, the misfit here is connected to the prior misfit from Equation \ref{eq:genCostFun} with measuring the weighted deviations of pore sizes and pore solidities with weights $\beta$ and $1-\beta$.
The KL divergence can be interpreted as a measure of information loss by approximation of the microstructure's descriptor with the surrogate's descriptor. 
The size and solidity of the pores are defined as in Equations \ref{eq:defSi}, \ref{eq:defSo}, where $P_{Si}^*$ corresponds to the dataset's size distribution function and $P_{Si}(\theta, \omega)$ is the surrogate microstructure's size distribution function. 
The definition for the solidities is analogous with $P_{So}^*$ and $P_{So}(\theta, \omega)$.
\footnote{Numerically it was found that using this cost function $J_{SS}$ leads to very few inclusions whose parameters follow the desired size and solidity distribution. However, this leads to a poor performance of $S_2$ correlation function. 
Hence a modified cost function was used of type $\hat{J}_{SS}(\theta, \omega)=J_{SS}(\theta, \omega)+\delta S_2[\theta, \omega]$, which alleviates this problem. }
\goodbreak
For the tuning of the Gaussian Random Field, the cost function $J_{SL}$ is used, which is based on the Two Point Correlation Function and Lineal Path function \ref{eq:defS2}, \ref{eq:defLp}
\begin{equation}
	J_{SL}(\theta, \omega)=\gamma\frac{\sum_{\boldsymbol{r}_i} \lvert S_2[\theta, \omega](\boldsymbol{r}_i)-S_2^*[\omega](\boldsymbol{r}_i)\lvert}{\sum_{\boldsymbol{r}_i} \lvert S_2^*(\boldsymbol{r}_i)\lvert}+(1-\gamma)\frac{\sum_{\boldsymbol{r}_i} \lvert L_p[\theta, \omega](\boldsymbol{r}_i)-L_p^*[\omega](\boldsymbol{r}_i)\lvert}{\sum_{\boldsymbol{r}_i} \lvert L_p^*(\boldsymbol{r}_i)\lvert}. 
\end{equation}
\goodbreak
As both problems are high dimensional and the computation of the cost functions is costly, an Efficient Global Optimization Method as proposed in \cite{jones_efficient_1998} is used in the implementation from \cite{saves_smt_2024}.

\begin{figure}[h]
\begin{algorithm}[H]
	\KwData{
        Dataset Microstructure $\chi_{GT}$
		Maximum Aspect Ratio $q_{max}$, 
	}
	\KwResult{Parameters $\theta$ for the Surrogate Generator}
\vspace{0.5cm}
	$\chi_{Train}\gets \chi_{GT}(ind(N, \phi_i))$ \tcp*{Sample Microstructure Slices with Even Vol. Frac. Distr.}
    $S_2^*, L_p^*, P_{Si}^*, P_{So}^*\gets \mathcal{I}_D(\chi_{Train})$\\
    $\alpha_0\gets 0$\\
    \While{$N<N_{ITER}, |C_{i+1}-C_i|>\delta$}{
		$\theta_{T}\gets EGO(J_{SS}^*[\theta_{GRF}])$        \\
        $\theta_{GRF}\gets EGO(J_{SL}^*[\theta_{T}])$        \\
	   $N\gets N+1$
    }    
\end{algorithm}
\caption{Optimization method to identify the parameter sets $\theta_T$ and $\theta_{GRF}$. 
The optimization is conducted for both parameter sets separately. In the first iteration the noise field is omitted by setting $\alpha=0$ from $\theta_{GRF}$ to zero. 
The efficient Global Optimization is implemented using \cite{saves_smt_2024} and takes the parameter set values of the other set from the previous iteration.}
\label{alg:stagScheme}
\end{figure}
In the first step, the topological support is optimized using $J_{SS}$ with no Gaussian random field noise. 
In the second step, given this topological support, the noise field is optimized. 
This staggered scheme is repeated with the updated noise field and noise level. 
This process, which is visualized in Algorithm \ref{alg:stagScheme}, is repeated and the final values are exported as the optimized parameters. 
\section{Mechanics Problem}
\label{sec:mechProb}
In order to compare the mechanical behavior of the surrogate microstructure and the dataset's microstructure, different mechanical problems were simulated using the FFT toolbox AMITEX FFT \cite{chen_fft_2019}. 
The computations were all conducted on slices from the dataset and the surrogate microstructure. 
The simulations were repeated for all the dataset slices, with the same number of tuned surrogate slices and the same volume fraction. 

\paragraph{Linear Elasticity}
An elastic homogenization procedure was conducted to predict deviations between the global, linear behavior of surrogate microstructure samples and samples from the dataset. 
This leads to the periodic boundary value problem of linear elasticity with peridicity vectors $\boldsymbol{L}_P$ following from the unit cell $\mathcal{D}$.
\begin{equation}
\begin{gathered}
		\text{div}(\boldsymbol{\sigma})=\text{div}\left(\mathbb{C}(\boldsymbol{x})[\boldsymbol{\varepsilon}]\right)=0, \boldsymbol{x} \in \mathcal{D}\\
		\mathbb{C}(\boldsymbol{x})=\mathbb{C}_{Mat}(1-\chi(\boldsymbol{x})), \boldsymbol{x} \in \mathcal{D}\\
		\boldsymbol{\varepsilon}=\frac{1}{2}\left(\nabla \boldsymbol{u} +\nabla^T \boldsymbol{u} \right),\		\left< \boldsymbol{\sigma} \right>_{\mathcal{D}}=\boldsymbol{\Sigma}\\		
		\boldsymbol{\varepsilon}\left(\boldsymbol{x}+\boldsymbol{L}_{P}\right)=\boldsymbol{\varepsilon}\left(\boldsymbol{x}\right),\ \boldsymbol{\sigma}\left(\boldsymbol{x}+\boldsymbol{L}_{P}\right)=\boldsymbol{\sigma}\left(\boldsymbol{x}\right),
\end{gathered}
\label{eq:LEBVP}
\end{equation}
with the average stress over the unit cell ($\left< \boldsymbol{\sigma} \right>_{\mathcal{D}}$) being equal to the prescribed stress $\boldsymbol{\Sigma}$. 
The elasticity tensor $\mathbb{C}$ is spatially dependent on the material parameters and the material identification function. 
\goodbreak
In order to compute Young's modulus and compression modulus in any direction $\boldsymbol{d}$, the compliance tensor needs to be determined. 
In the 2D case for the slices, 3 parameters need to be determined by computing the three separate elastic loadings ($\Sigma_{11}$, $\Sigma_{22}$, $\Sigma_{12}$), where the $\Sigma_{ij}$ correspond to the homogenized stresses.

Following \cite{bohlke_graphical_2001}, the directional Young's modulus and compression modulus can be written as 
\begin{align}
	E(\boldsymbol{d})=\left(\boldsymbol{d} \otimes \boldsymbol{d} \cdot \mathbb{S} \left[ \boldsymbol{d} \otimes \boldsymbol{d} \right] \right)^{-1} &&
	K(\boldsymbol{d})=3\left(\boldsymbol{I} \cdot \mathbb{S} \left[ \boldsymbol{d} \otimes \boldsymbol{d} \right] \right)^{-1} 
\end{align}
in order to visualize them in polar plots. 

\begin{figure}[H]
	\centering
	\includegraphics{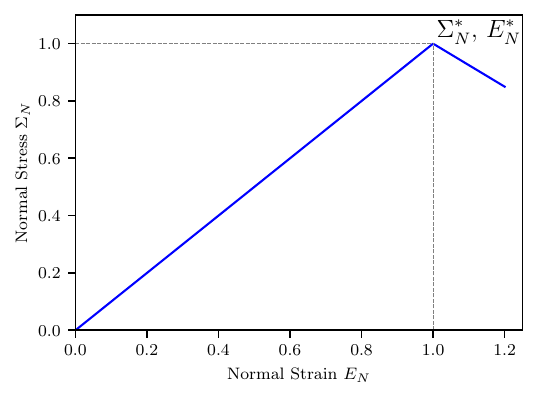}
	\caption{Illustration of Peak Homogenized Stress and Strain notion. The peak homogenized stress $\Sigma_N^*$ is defined as the largest normal stress during the normal loading. The peak homogenized strain $E_N^*$ corresponds to the normal strain at this point.}
	\label{fig:defPeakStressStrain}
\end{figure}

\paragraph{Elastic Brittle Fracture}
To predict deviations in the failure behaviour of surrogate samples, an elastic brittle fracture failure model was computed for slices of the dataset microstructure and the surrogate samples.  
The computation of the critical load accounting for brittle fracture failure has been conducted using a phase field model as proposed in \cite{miehe_thermodynamically_2010, miehe_phase_2010} with the implementation from \cite{chen_fft_2019}. 
Here, the crack path is predicted using a variational formulation, which is an extension to equation \ref{eq:LEBVP}, as the constitutive law will depend on the phase damage variable d.
\begin{equation}
	\begin{gathered}
		\text{div}\left(\boldsymbol{\sigma}(\boldsymbol{u}, d)\right)=0\\
		\boldsymbol{\sigma}=g(d)\left[\lambda \left< \text{tr}(\boldsymbol{\varepsilon}) \right>_+\boldsymbol{I}+2\mu \boldsymbol{\varepsilon}_+\right]+\left[\lambda \left< \text{tr}(\boldsymbol{\varepsilon}) \right>_-\boldsymbol{I}+2\mu \boldsymbol{\varepsilon}_-\right]\\
		\lambda(\boldsymbol{x})=\lambda(1-\chi(\boldsymbol{x})), \hspace{1cm}\boldsymbol{x} \in \mathcal{D}\\
		\mu(\boldsymbol{x})=\mu(1-\chi(\boldsymbol{x})), \hspace{1cm}\boldsymbol{x} \in \mathcal{D}\\		
		g_c(\boldsymbol{x})=g_c, \hspace{1cm} \boldsymbol{x} \in \mathcal{D}\\
		\frac{g_c}{l_c}\left[d-l_c^2\Delta d\right]=2(1-d)\mathcal{H}(\boldsymbol{\varepsilon}), \\
		<x>_{\pm}=(x\pm\lvert x \lvert)/2, \ \boldsymbol{\varepsilon}^\pm=\sum_{i=1}^{3}< \varepsilon^i>_\pm \boldsymbol{n}^i \otimes \boldsymbol{n}^i 
		\end{gathered}
\end{equation}
where $g(d)$ describes the degradation through damage and is chosen as $g(d)=(1-d)^2+k$. 
$\boldsymbol{\varepsilon_\pm}$ corresponds to a spectral split with $\varepsilon^i$ and $\boldsymbol{n}^i$ as eigenvalues and eigenvectors. 
The computations are again conducted using AMITEX based on \cite{chen_fft_2019}. 
\goodbreak
The peak normal stress and corresponding strain are defined as in \cite{alessi_phenomenological_2018} following the definition from \cite{marigo_overview_2016}. 
The peak normal stress is defined as the maximum stress that occurs during the strain based loading and the peak normal strain is defined as the normal strain at this point. 
This procedure is visualized in Figure \ref{fig:defPeakStressStrain}.

\section{Application of Surrogate Microstructure Generation for Tristructural Iso\-tropic Nuclear Fuel Particles}
\label{sec:TrisoProb}
\begin{figure}[h]
	\centering
	\begin{subfigure}{0.56\linewidth}
		\centering
		\includegraphics[width=0.7\linewidth]{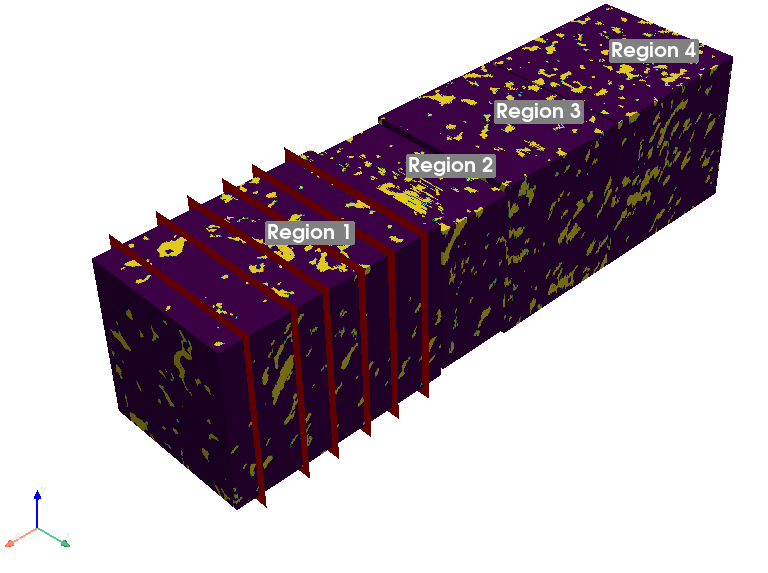}
		\caption{Dataset with visualization of slices taken from region 1 (red). The slices are computed based on Latin Hypercube Sampling to ensure a diverse spread of data}
		\label{fig:dsRegs}
	\end{subfigure}
	\begin{subfigure}{0.43\linewidth}
		\centering
		\includegraphics[cframe=pythonBlue 2pt, width=0.8\linewidth]{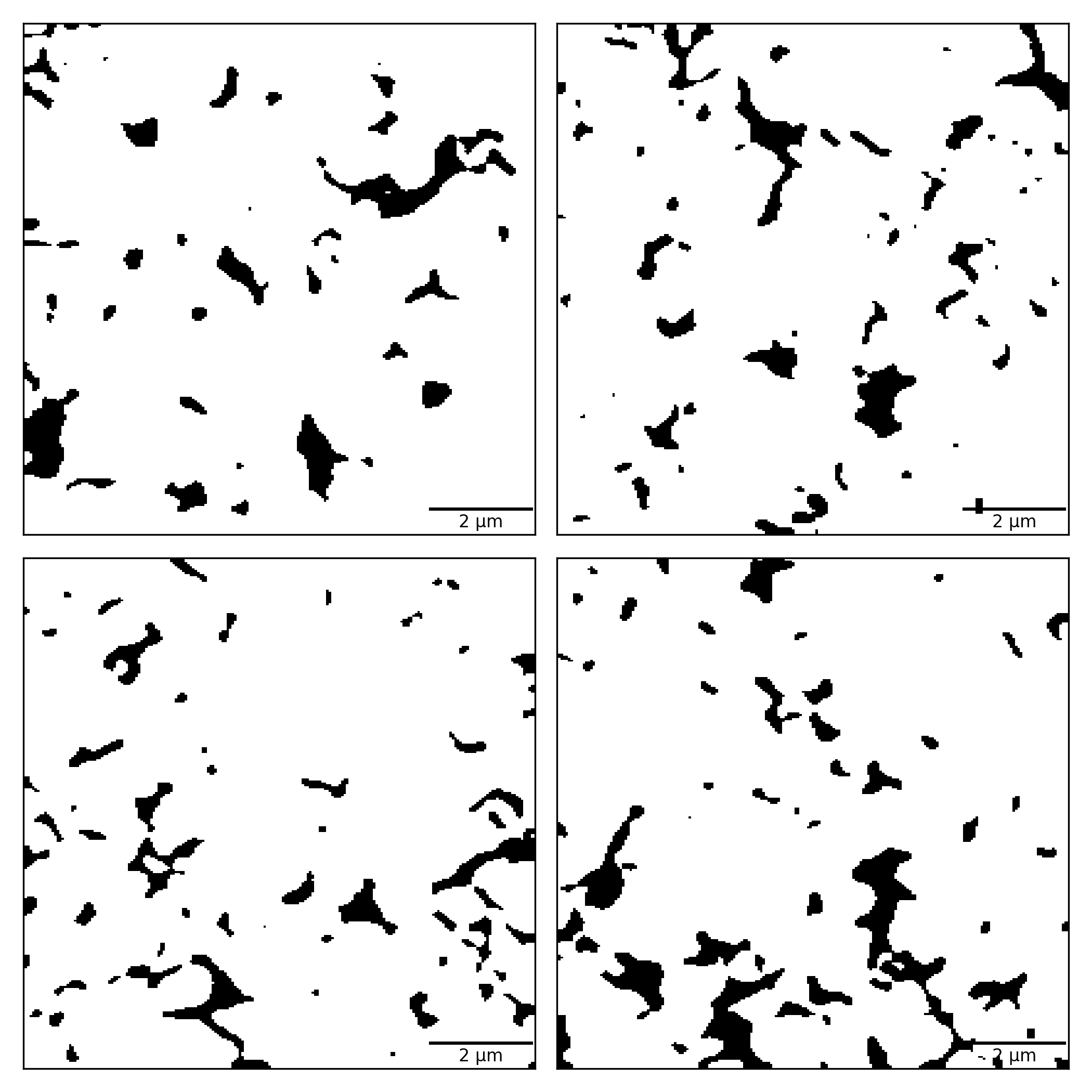}
		\caption{Radial slices from region 1 of the dataset}
		\label{fig:dsSlices}
	\end{subfigure}
	\caption{Voxelized Tristructural Isotropic Buffer dataset from \cite{griesbach_microstructural_2023}. The left is on the inner side of the buffer layer.}
\end{figure}

\paragraph{Data Curation}
To generate buffer layer surrogate samples with optimized microstructural properties, the parameters defined in section \ref{sec:surMSGenId} are optimized to minimize the misfit to dataset slices along the radial direction from \cite{griesbach_microstructural_2023}.
As part of the work in \cite{griesbach_microstructural_2023}, the microstructure of the buffer layer of TRISO fuel particles was analyzed using focused ion beam (FIB)–scanning electron microscope (SEM) tomography. 
The microstructural slices were grouped in 8 sections by position, which can be seen in figure \ref{fig:dsRegs}, where the first four groups are plotted. 
As the segmentation was conducted for individual pores based on a machine learning approach using Dragonfly 2021.1, a mesh of the pores per region is given. 
To compute the image descriptors used, a voxelized version on specific slices was conducted. 
The voxelization was conducted with a python script based on the vtk toolbox \cite{schroeder_visualization_2006}

\paragraph{Mechanical Comparison}
In order to assess the performance of the surrogate microstructure generator, the mechanical problems introduced in section \ref{sec:mechProb} are computed for both surrogate slices and dataset slices. 
For the elastic part, the elastic constants of slices from the dataset and the surrogate microstructure are compared. \\
For the brittle fracture problem, the peak homogenized stress and strain is compared, which is visualized in Figure \ref{fig:defPeakStressStrain}.
During loading, the normal strain is increased linearly from $0.0$ to $0.05$. 
If the peak normal stress was found before a strain of $0.0475$, the specimen is assumed to have failed.
If the specimen has not failed up to this point, the simulation was repeated up to a mean strain of $0.075$ with the same simulation parameters. 
\goodbreak
For region 1, 564 slices were computed along the radial direction with a voxel size of 19x25 nm and a tickness of 25nm. 
Based on this, voxelized slices and the descriptors were computed following method elaborated in Section \ref{sec:surMSGenId}. 
The mechanical properties of the TRISO buffer layer microstructure are chosen according to the data for nuclear grade graphite in \cite{chakraborty_phase-field_2016, burchell_modeling_2007} for the elastic and brittle fracture properties.

\begin{table}[h]
	\centering
	\begin{minipage}{0.45\textwidth}
		\centering
		\begin{tabular}{l|rl}
			\toprule
			$l_{Voxel}$ & 12.5 & nm \\
			Length Scale & 30 & nm \\
			$N_{\text{slices}}^{\text{GT}}$& 564 & - \\
			$N_{\text{slices}}^{\text{Surr.}}$& 564 & - \\
			\bottomrule
		\end{tabular}
	\end{minipage}
	\hfill
	\begin{minipage}{0.45\textwidth}
		\centering
		\begin{tabular}{l|rl}
			\toprule
			Lamé Constant, $\lambda$ & 1.707 & GPa\\
			Shear Modulus, $\mu$ & 3.034 & GPa \\
			Griffith Criterion, $g_c$ & 10.68 & MPa$\mu$ m  \\
			\bottomrule
		\end{tabular}
	\end{minipage}
	\caption{Length scale parameters for the extracted material images and material parameters for the simulation of surrogate and dataset slices from \cite{chakraborty_phase-field_2016}}
	\label{tab:simMatParas}
\end{table}
\section{Results}
\label{sec:res}

\subsection{Image Analysis}
\begin{figure}[h]
	\centering
	\begin{subfigure}[t]{0.49\linewidth}
		\centering
        \includegraphics[cframe=pythonBlue 2pt, width=\linewidth]{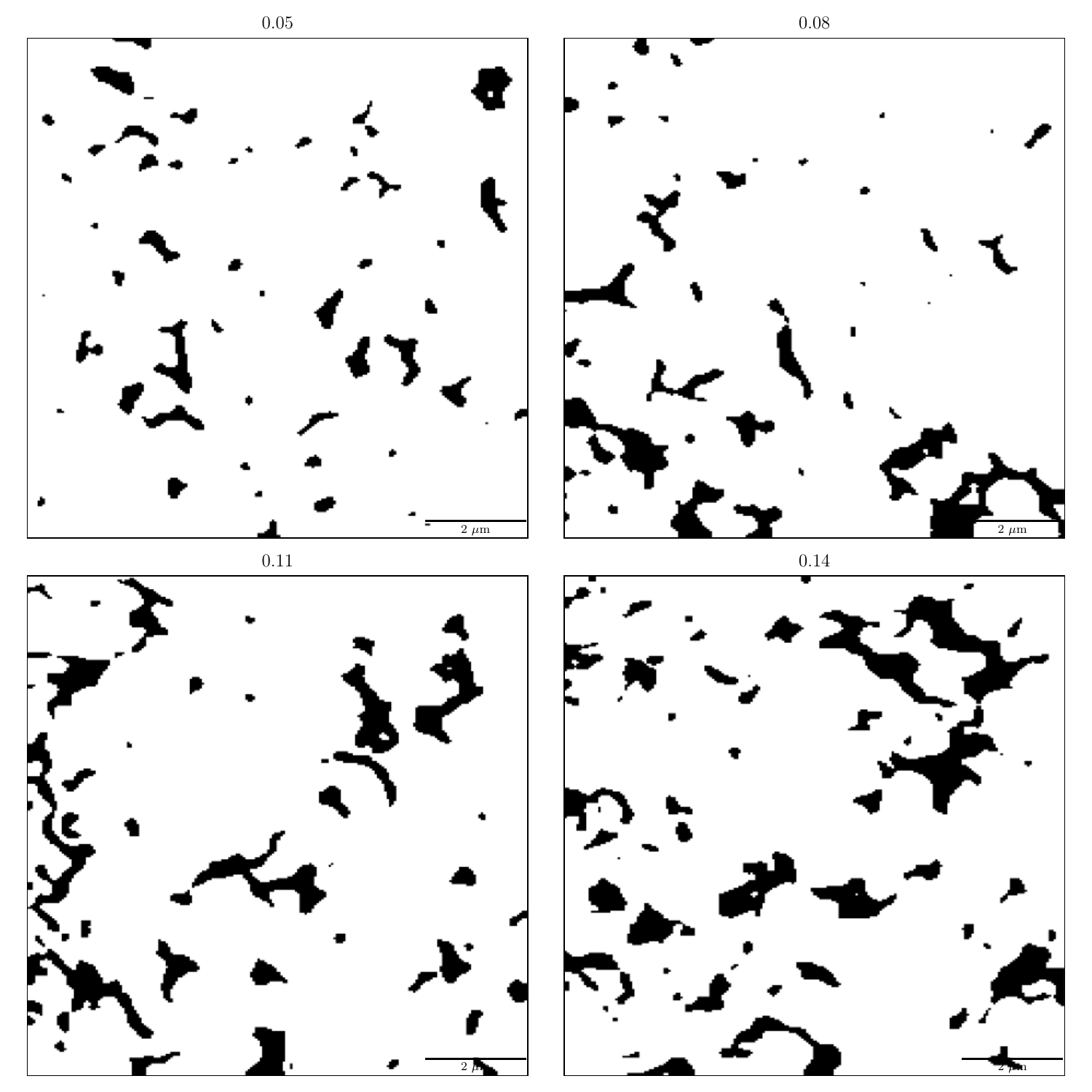}
        \caption{Radial microstructure slices from the dataset (slices to cover a wide range of volume fractions)}
		\label{fig:resMS_gt}
	\end{subfigure}
	\begin{subfigure}[t]{0.49\linewidth}
		\centering
		\includegraphics[cframe=pythonOrange 2pt, width=\linewidth]{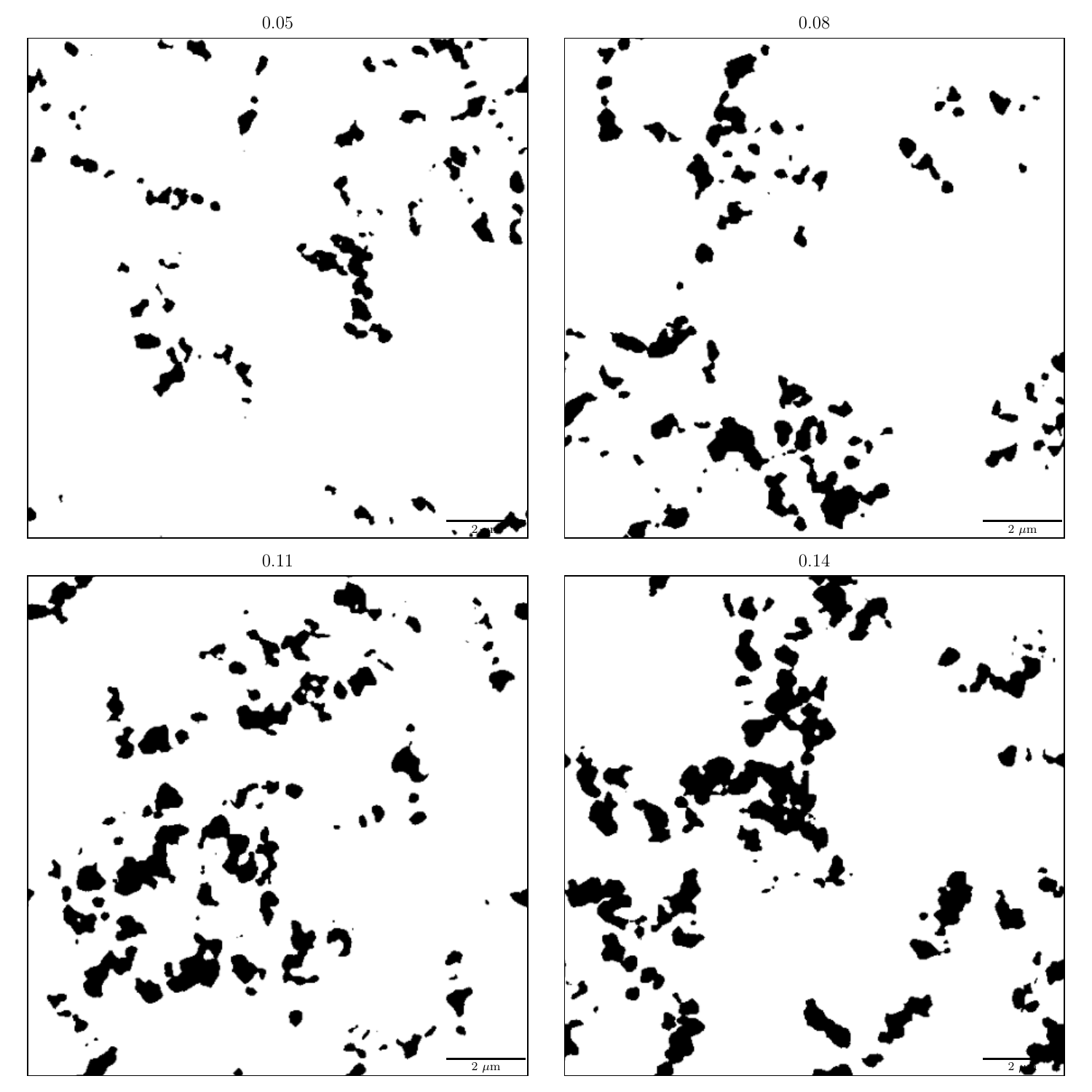}
		\caption{Samples from tuned surrogate microstructure generator with equivalent volume fractions as the dataset on the left.}
		\label{fig:resMS_surr}
	\end{subfigure}

	\caption{Comparison of samples from surrogate and dataset microstructure of equivalent volume fractions}
	\label{fig:resMS}
\end{figure}
\paragraph{Microstructure Morphology}
Figure \ref{fig:resMS} presents slices from the dataset microstructure (\ref{fig:resMS_gt}) and surrogate microstructure (\ref{fig:resMS_surr}) with equivalent volume fractions (5, 8, 11 and 14 \%). 

A visual inspection of the morphologies shows that the size of the inclusions (black) is well represented. 
However, the surrogate slices show regions with very low porosities that are less present in the dataset slices. 
This is particularly dominant in the top left slice, which has very few pores on the left and can be explained by the topological support and the clustering, where the clustering leads to an accumulation of pores in a region and a lack of pores in another region. 
Furthermore, pores from the dataset microstructure frequently show sharp edges that are not observed in the surrogate microstructure. 
\begin{figure}[H]
	\centering
	\begin{subfigure}[b]{0.9\textwidth}
		\centering
		\includegraphics[width=\linewidth]{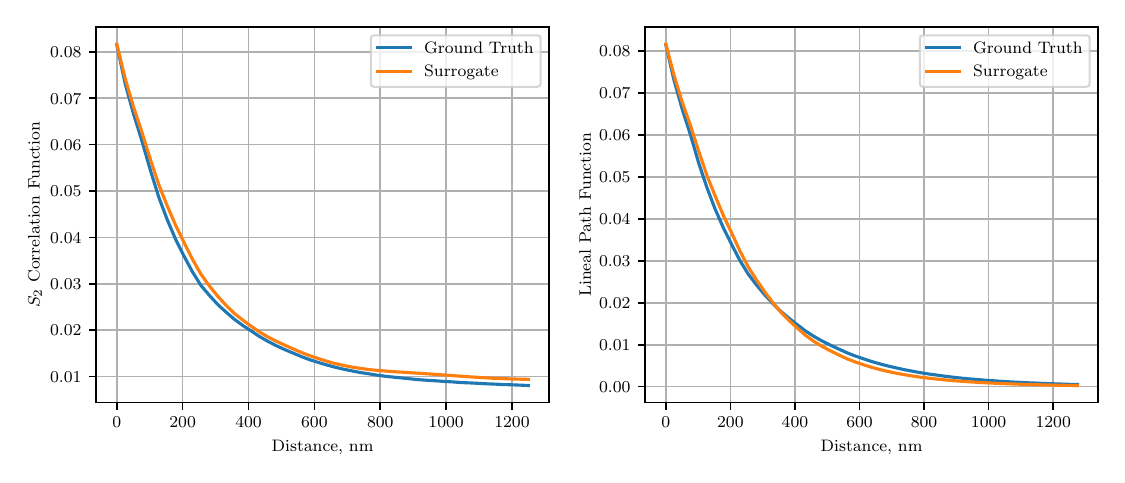}
		\caption{Comparison of $S_2$ correlation function and lineal path function from the dataset and surrogate microstructure.}
		\label{fig:descS2Lp}
	\end{subfigure}
	\begin{subfigure}{0.45 \linewidth}
		\centering
		\includegraphics[width=\linewidth]{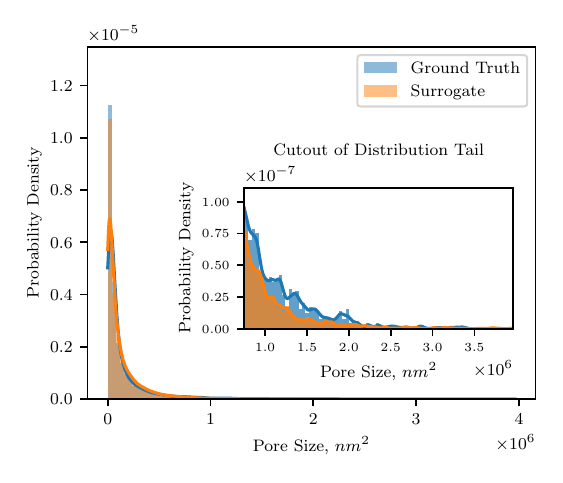}
		\caption{Comparison of the size distribution functions for pores from 50 slices from Ground Truth dataset and surrogate generator}
		\label{fig:descSi}
	\end{subfigure}
	\begin{subfigure}{0.45 \linewidth}
		\centering
		\includegraphics[width=\linewidth]{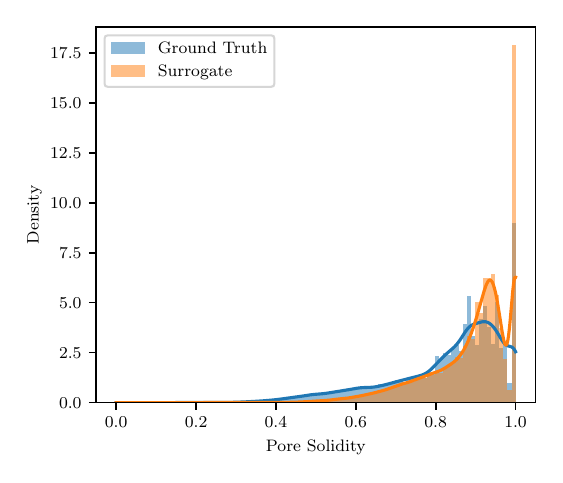}
		\caption{Comparison of the Solidities distribution functions of pores from 50 slices from Ground Truth dataset and surrogate generator}
		\label{fig:descSo}
	\end{subfigure}
	\caption{Comparison of Resulting Image Descriptors between slices from region 1 of the dataset and the tuned surrogate generator.}
	\label{fig:resID}
\end{figure}
\paragraph{Image Descriptors}
Figure \ref{fig:resID} represents the Two-Point correlation function and lineal path function that were used for the optimization of the surrogate generator. 
The $S_2$ correlation function and lineal path function (see \ref{fig:descS2Lp}) get matched very well at low separations. 
At higher distances (greater 200 nm), the $S_2$ correlation function of the surrogate microstructure includes an overestimation the correlation between points. 
This, again, can be attributed to the clustering of pores in regions leading to an overestimation of the correlation function in these regions. 
In figure \ref{fig:descSi} and \ref{fig:descSo}, the size and solidity distribution function of pores from the dataset and surrogate generator are displayed. 
Furthermore, a Gaussian kernel density estimation is included. 
The pore size distributions show a good representation of the dataset pore sizes in the surrogate microstructure pore sizes. 
However, the dataset microstructure has fewer very small pores but slightly more very large pores (see the cutout of the tail of the distribution). 
The solidity distribution of pores from the dataset and from the surrogate generator shows that the solidities in a range between $0.7$ to 0.98 get approximated well. 
However, the surrogate microstructure slices have very few pores with a solidity below 0.7 and a density that is too high for perfectly convex pores (solidity of 1). 

\subsection{Mechanical Problems}
\begin{figure}[H]
	\centering
	\includegraphics[width=\linewidth]{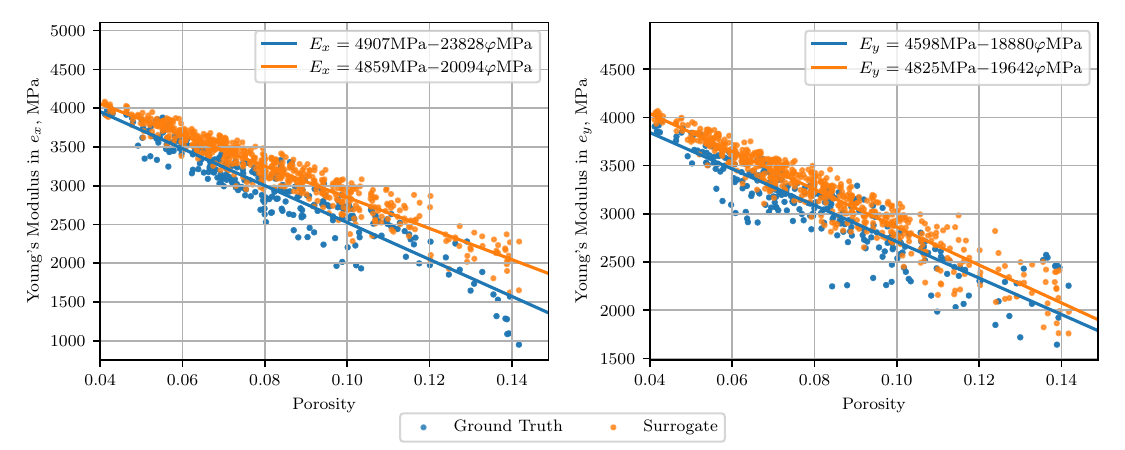}
	\caption{Comparison of the elastic moduli in $\boldsymbol{e}_x$ (left) and $\boldsymbol{e}_y$ (right) at different porosities of the surrogate and dataset slices.}
	\label{fig:resPorYM}
\end{figure}

\begin{figure}[h]
	\centering
	\includegraphics[width=\linewidth]{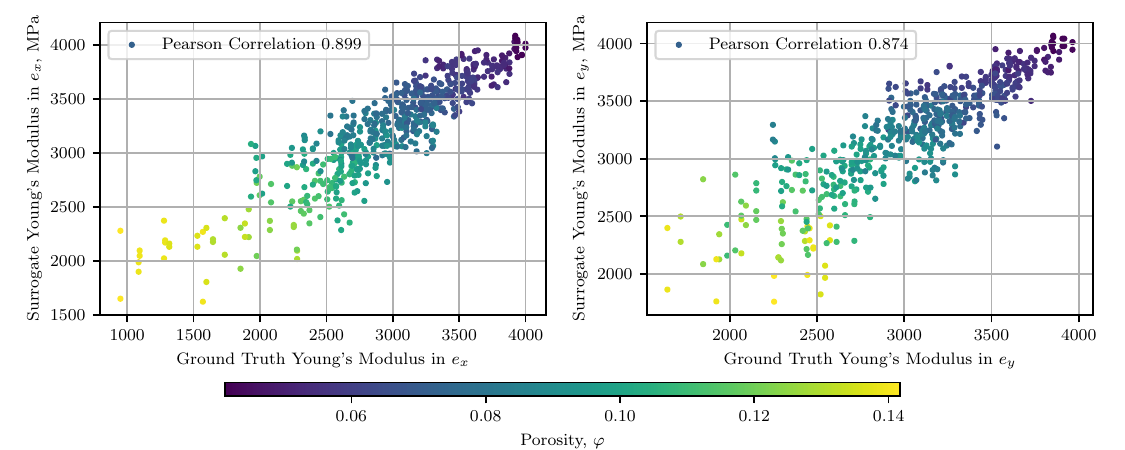}
	\caption{Comparison of the elastic moduli in $\boldsymbol{e}_x$ (left) and $\boldsymbol{e}_y$ (right), with dataset slices on the abscissa and surrogate slices on the ordinate, at different porosities (color encoded).}	\label{fig:ElasticModuliDetermination}
\end{figure}
\begin{figure}[H]
	\centering
	\includegraphics[width=0.78\linewidth]{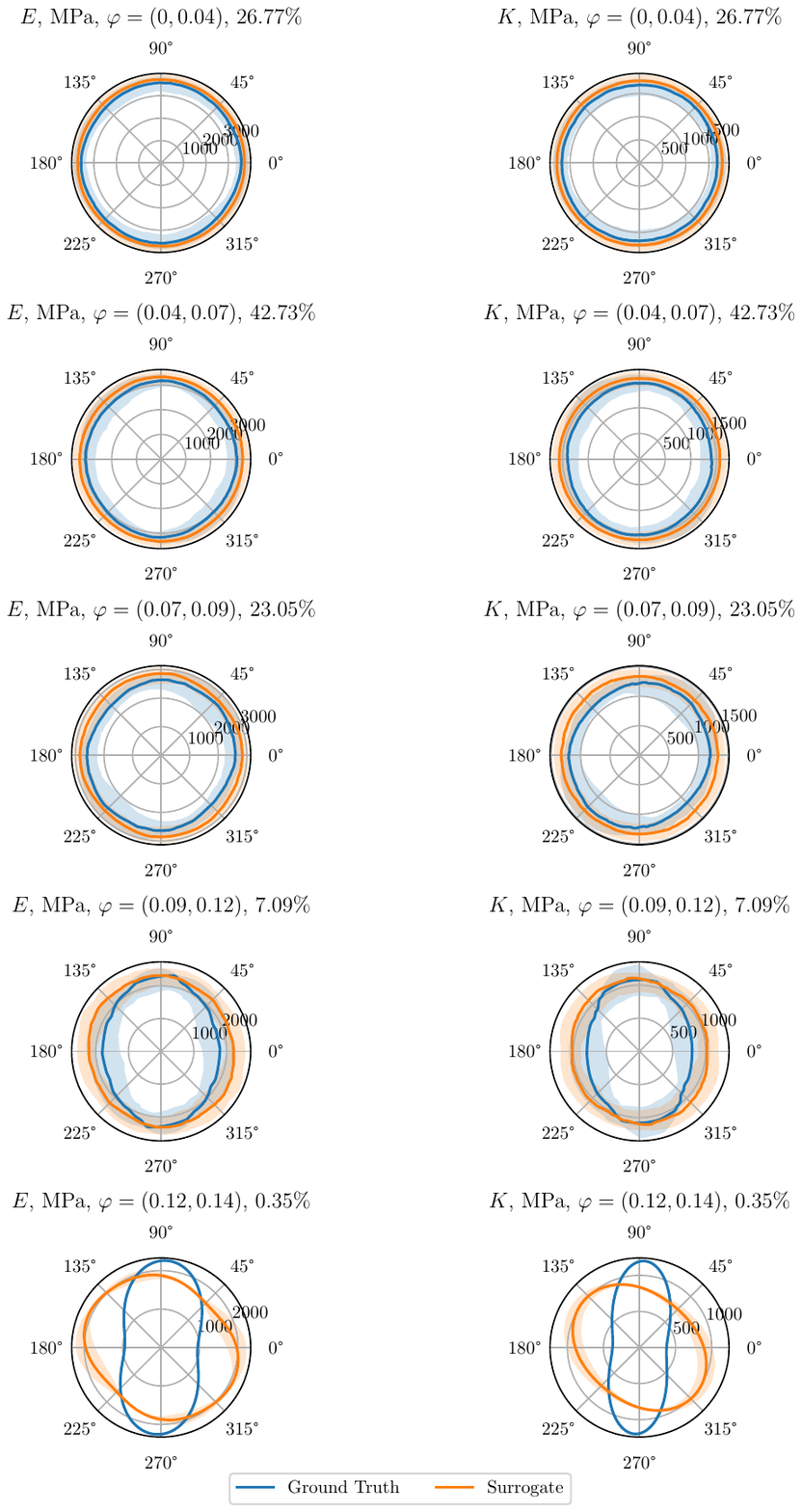}
	\caption{Directional Homogenzation of the surrogate and dataset microstructure for different volume fractions. Each time, the 80\% interval is shown as shading. 
    At porosities over 9\%, anisotropic effects gain importance. This explains the increased deviations at higher porosities as they are not accounted for in the surrogate microstructure.}
	\label{fig:anisoYM}
\end{figure}

\paragraph{Homogenized Elastic Moduli}
Figure \ref{fig:resPorYM} shows the development of the homogenized Young's moduli in directions $\boldsymbol{e}_x$ and  $\boldsymbol{e}_y$ and a linear interpolation of the development over the porosities for the dataset and surrogate slices. 
At low porosities, the difference between the surrogate elastic moduli and dataset elastic moduli is low, particularly in $\boldsymbol{e}_y$. 
However, at higher porosities, the deviation increases, particularly in $\boldsymbol{e}_x$. 
This difference can be seen in the linear interpolation of the correlation of the elastic moduli with porosities. 
Whilst the deviation of the interpolations' slopes is below 1\% in $\boldsymbol{e}_y$, the deviation in $\boldsymbol{e}_x$ is around 20\%. 
This can partially be attributed to the few dataset slices at higher porosities. 
Hence, the features at low porosities are matched better. 
Furthermore, the morphology of slices with very high porosity is more complex, as can be seen in \ref{fig:resMS_gt}. 
\goodbreak
Figure \ref{fig:ElasticModuliDetermination} shows the relationship between the surrogate and dataset Young's modulus in $\boldsymbol{e}_x$ and $\boldsymbol{e}_y$ direction. 
The spread in surrogate Young's moduli for equivalent dataset Young's modulus is larger in $\boldsymbol{e}_y$ direction, which is especially dominant at lower Young's moduli. 
This can be seen in the lower Pearson correlation between the surrogate and dataset stiffnesses. 
Nevertheless, the error between the surrogate and dataset stiffness remains is lower at lower stiffnesses, which is consistent with the findings Figure \ref{fig:anisoYM}.
\goodbreak
Figure \ref{fig:anisoYM} shows the directional Young's moduli of both the surrogate samples and the dataset samples for different volume fraction ranges with the 10th percentile, median, and 90th percentile. 
At low porosities, the directional elastic moduli get approximated well. However, the overestimation of the stiffnesses increases with increase in volume fraction, which is consistent with the previous analysis of figure \ref{fig:resPorYM}. 
Another feature is the anisotropy of elastic moduli for porosities over around 10 \% . 
This anisotropy and stronger direction $\boldsymbol{e}_y$ does not get captured in the surrogate model as all the parameters of the generator are assumed to be isotropic. 
This, however, only relates to around 7.5\% of the microstructure slices. 
\begin{figure}[H]
	\centering
	\begin{subfigure}{0.8\linewidth}
		\centering
		\includegraphics[width=\linewidth]{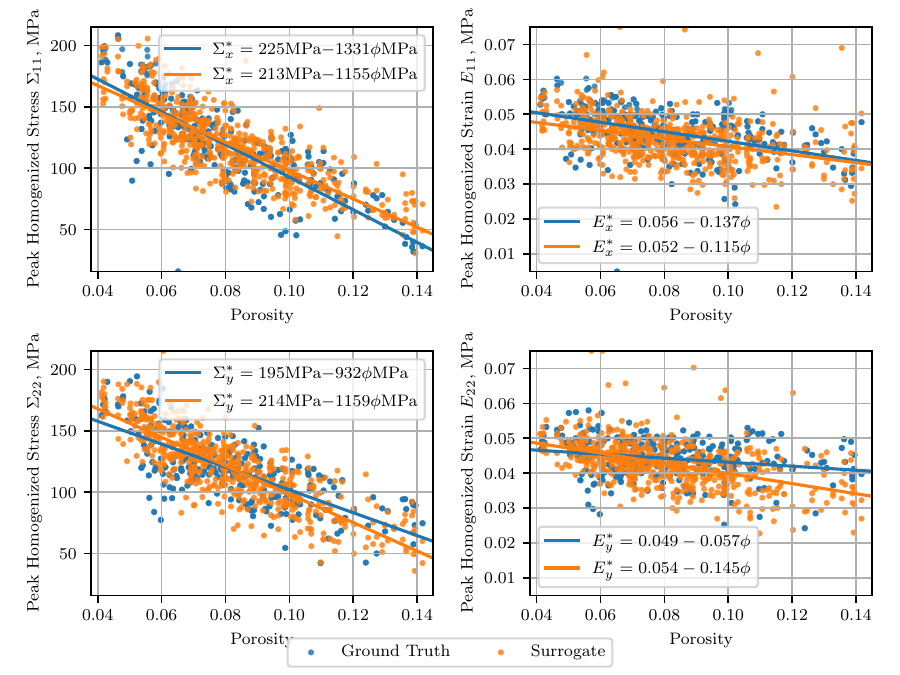}
		\caption{Development of the brittle fracture critical peak homogenized stress and critical strain at different porosities}
		\label{fig:BFPor}
	\end{subfigure}
	\begin{subfigure}{0.8\linewidth}
		\centering
		\includegraphics[width=\linewidth]{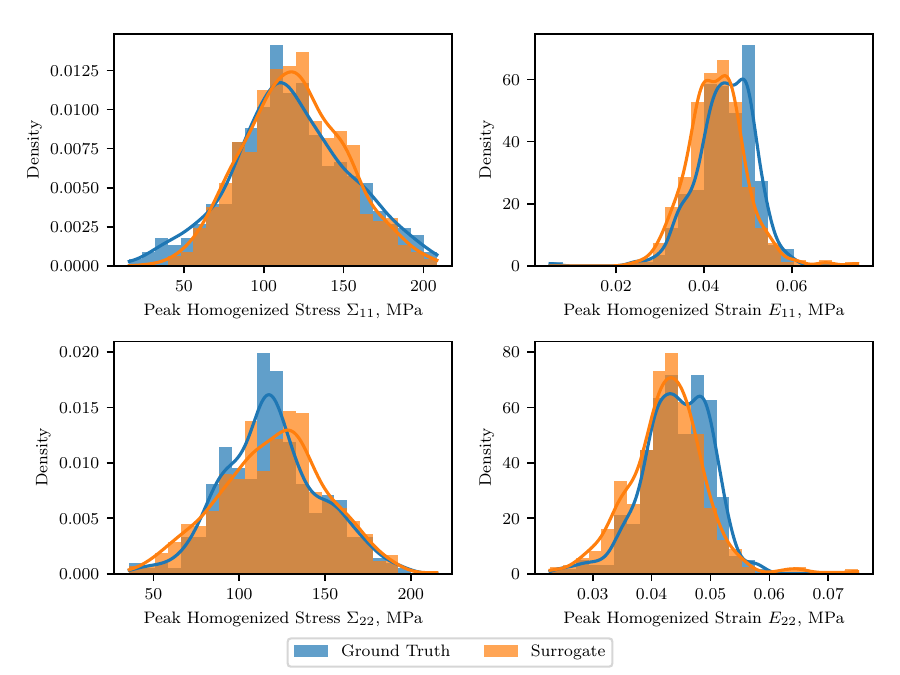}
		\caption{Histogram of the critical strains and stresses of brittle fracture.}
		\label{fig:histBFCV}
	\end{subfigure}
	\caption{Distribution of brittle fracture critical stresses and strains for the surrogate microstructures and the dataset microstructures.}
\end{figure}
\paragraph{Elastic-Brittle Fracture Critical Values}
Figure \ref{fig:BFPor} shows the degradation of critical stress and critical strain with porosities in both $\boldsymbol{e}_x$ and $\boldsymbol{e}_y$ with a linear approximation. 
It can be seen that the critical stress decreases with an increasing porosity whilst the critical strain remains relatively constant (see the slope of the linear interpolation of the critical strain values). 
Figure \ref{fig:histBFCV} shows a histogram of the critical stress and strain brittle fracture values. 
Here, a very good agreement was found, with the medians showing deviation a deviation of 2.9\% for $\Sigma_{22}^*$ and 4.2\% for $\Sigma_{11}^*$, again, the surrogate slices were found to be tougher and the deviations were larger in $\boldsymbol{e}_x$. 
The information to be extracted from figure \ref{fig:BFPor} is limited as the critical values show a very high spread for the same porosities. 
\section{Discussions}
\label{sec:concs}
The application of the developed surrogate generation approach to buffer layer microstructures from tristructural isotropic nuclear fuel particles show the viability of the surrogate approach. 
The generated surrogate samples can be applied to the PARFUME code to model more detailed the microstructures. 
Here, the sampling of the homogenized mechanical properties for the simulation of the buffer layer can be conducted using the proposed method the performance of the fuel kernel. 
This workflow allows the prediction of the performance using few buffer layer samples. 
Hence, the viability of changes in production can be evaluated at an early stage and subsequent adpations in the production can be made.
Due to the sampling capabilities of the proposed method, failure probabilities under given loading conditions can be computed with higher fidelity. 
\goodbreak
To enhance the current results, three dimensional simulations should be conducted including thermomechanical coupling, which can include more complex boundary conditions. 
At higher porosities of the slices, the expected error in the predicted mechanical properties with respect to the mechanical properties increases. 
As this is mainly due to the increased effect of anisotropy, anisotropic effects should be accounted for in future versions of the topological support. 
\goodbreak
Furthermore, the developed surrogate model should be adapted to include modeling of the morphological developments based on the radiation through the comparison of the morphology of the buffer layer from pre and post-irradiation experiments.
The use of surrogate microstructure generator allows to add a model of the development of the morphology in this process. 
This means that the mechanical simulations can be conducted at different steps of irradiation degradation to model and extrapolate the damage behavior over relevant timescales. 
\goodbreak
Perspectively, the surrogate microstructure generator for the buffer layer should be coupled with surrogate microstructure generators for other regions. 
Then, a complete surrogate model can be developed for TRISO nuclear fuel particles. 

\section*{Acknowledgements}
The authors gratefully acknowledge Thevamaran Ramathasan, Yongfeng Zhang, and Claire Griesbach , Department of Mechanical Engineering, University of Wisconsin-Madison, for their insightful discussions and valuable input throughout the development of this work. We also thank them for their help working with their dataset, which was instrumental in our study.\\
This work was granted access to the high-performance computing resources of the IDCS support unit from Ecole Polytechnique.
\newpage

\bibliographystyle{elsarticle-num} 
\bibliography{references}

\newpage
\appendix
\section{Parameters of the Surrogate Material}
\paragraph{Parameter Description} To recapitulate, the following parameters have been introduced as part of the parameter vector $\theta=[\theta_{TS}, \theta_{GRF}]$. 
The given values correspond to the values obtained through the optimization. 
\begin{itemize}
	\item Topological Support, $\theta_{TS}$
	\begin{itemize}
		\item Number of Clusters $N_c=66$
		\item Binomial Success $p_c=0.5$
		\item Number of Ellipsoids $N_e=182$
		\item Random Walk Stepsize $s=0.89\mu$m
		\item Angle Standard Deviation $\alpha_d=0.1$
		\item Relative Particle Radius $d=0.01$
		\item Maximum aspect ratio $q_{max}=1.5$
		\item Morphological Disk size $D=140$nm
	\end{itemize}
	\item Gaussian Random Field, $\theta_{GRF}$
	\begin{itemize}
		\item Correlation Length, $l=0.225\mu$m 
		\item Smoothness $\nu=5/2$
		\item Noise level $\alpha=0.5$
	\end{itemize}
\end{itemize}
\end{document}